\documentclass[aps,pre,preprint,eqsecnum,amssymb]{revtex4}
\usepackage{graphicx,color,psfrag}
\usepackage{amsmath}
\usepackage{bm}
\begin{document}
\title{Heat currents in qubit systems}
\author{Hans C. Fogedby}
\email{fogedby@phys.au.dk,hans.fogedby@gmail.com}
\affiliation{Department of Physics and Astronomy, 
University of Aarhus, Ny Munkegade,
8000 Aarhus C,
Denmark}
\begin{abstract}
There is a current interest in quantum thermodynamics in the context of open quantum 
systems. An important issue is the consistency of quantum thermodynamics, in particular
the second law of thermodynamics, i.e., the flow of heat from a hot reservoir to a cold
reservoir. Here recent emphasis has been on composite system and in particular the issue
regarding the application of local or global master equations. In order to contribute to this
discussion we discuss two cases, namely as an example a single qubit and as a simple 
composite system two coupled qubits driven by two heat reservoirs at different 
temperatures, respectively. Applying a global Lindblad master equation approach 
we present explicit expressions for the heat currents in agreement with the second law 
of thermodynamics. The analysis is carried out in the Born-Markov approximation. 
We also discuss issues regarding the possible presence of coherences in the steady state.
\end{abstract}
\maketitle
\section{Introduction}
There is a current interest in quantum thermodynamics
\cite{Alicki79,Barra15,Kosloff13a,Kosloff14,Mari12,Kosloff19,
Levy20,Colla22,Linden10}.
Thermodynamics is a universal theory
and must from a fundamental point of view emerge from a microscopic point of view.
In this context the emergence of in particular the second law of thermodynamics. 
In the Clausius formulation the second law implies that heat flows from a hot reservoir 
to a cold reservoir and this essential property should also hold in the quantum regime. 
In non equilibrium quantum statistical mechanics the instantaneous heat current is 
subject to both statistical and quantum fluctuations, however, the mean heat 
current must obey the second law of thermodynamics.

The natural framework for discussing these issues is the theory of open quantum systems
\cite{Breuer06,Rivas12,Davies76}.
Here the focus is on a small controllable quantum system interacting with an environment 
which can be either radiation fields or heat baths. In order to obtain a steady heat current
we must drive the quantum system by several heat reservoirs, typically two reservoirs, 
maintained at different temperatures. 

Whereas the case of a single qubit (two-level atomic system) driven by two heat bath 
is easily analysed \cite{Aurell19} and discussed in the present paper, the discussion in the 
literature is usually focussed on composite systems 
\cite{Kohler05,Kosloff14}.
Here a recent discussion has focussed
on the connection between a global master equation approach as applied in the
present paper and a local master equation approach in which the interactions in the 
composite system are ignored in the dissipative kernels.
It has been shown that in the 
case of a  local master equation approach the apparent violation of the second law 
can be remedied by using a protocol of repeated interactions; this scheme accounts 
for a work contribution which together with the heat contribution restores the 
validity of the second law 
\cite{Barra15,Pereira18,DeChiara18,Hewgill21}.
In the case of coupled quantum oscillators driven by two 
heat reservoirs a global approach was discussed in  \cite{Levy14} resolving the 
issue regarding the second law.
We also note recent work \cite{Soret22} dealing with the general issues regarding the 
consistency of quantum master equations in the analysis of quantum thermodynamics.

In the present work, motivated by the study in
\cite{Levy14}, we discuss a composite system of two coupled qubits driven
by two heat reservoirs. Coupled qubit systems has been studied extensively 
but mainly in the context of quantum entanglement
\cite{delValle10,Liao11,Fischer13,Duan13,Paneru20,Schlosshauer19,Aolita15,
Decordi17,Bresque21,Hewgill18,Hu18,Karimi17};
in the present paper, however, the focus 
is on quantum thermodynamics and the second law of thermodynamics.

To the best of our knowledge explicit expressions for the heat currents in a coupled two-qubit
systems exhibiting consistency with the second law of thermodynamics are,
however, not available. We therefore list our findings below. In the simple case
of a single qubit driven by two reservoirs we find for the currents $q^A$ from reservoir
$A$ to the qubit the expression
\begin{eqnarray}
&&q^A=
\frac{g_Ag_B\omega_0}
{g_A(1+2n_A)+
g_B(1+2n_B)}[n_A-n_B],
\label{qA}
\end{eqnarray}
where $\omega_0$ is the energy splitting, $g_{A,B}\equiv g_{A,B}(\omega_0)$ the
spectral functions, and $n_{A,B}\equiv 1/(\exp(\omega_0/T_{A,B})-1)$ the Planck
distributions.

In the case of coupled qubits we obtain
\begin{eqnarray}
&&q^A=q^A_++q^A_-,
\label{2qA}
\\
&&q^A_+=\frac{(\alpha\beta)^2g_A(\omega_+)g_B(\omega_+)\omega_+}
{\alpha^2g_A(\omega_+)(1+2n_A(\omega_+))+
\beta^2g_B(\omega_+)(1+2n_B(\omega_+))}
[n_A(\omega_+)-n_B(\omega_+)],
\label{2qAp}
\\
&&q^A_-=\frac{(\alpha\beta)^2g_A(\omega_-)g_B(\omega_-)\omega_-}
{\beta^2g_A(\omega_-)(1+2n_A(\omega_-))+
\alpha^2g_B(\omega_-)(1+2n_B(\omega_-))}
[n_A(\omega_-)-n_B(\omega_-)],
\label{2qAm}
\end{eqnarray}
where 
$\omega_\pm=(\omega_1+\omega_2)/2\pm\sqrt{(\omega_1-\omega_2)^2/4+\lambda^2}$
are the perturbed levels, $\omega_{1,2}$ the unperturbed levels, $\lambda$ the qubit
coupling, and 
$\alpha=\cos\theta, \beta=\sin\theta, \tan\theta=2\lambda/(\omega_1-\omega_2)$.
The expressions (\ref{qA}-\ref{2qAm}) exhibit a similar structure and are by inspection consistent
with the second law.

In discussing open quantum systems described by a master equation
the focus is on the density matrix $\rho_{pp'}(t)$, where the index $p$ refers to an 
energy basis. Here we  distinguish between populations characterised by the 
diagonal density matrix elements $\rho_{nn}(t)$ and coherences described by 
the off-diagonal matrix elements $\rho_{nm}(t)$, $ n\neq m$. For physical reasons
the populations must be positive, $\rho_{nn}\ge 0$ and, moreover, satisfy the trace 
condition $\sum_n\rho_{nn}=1$ (conservation of probability). Such a requirement
does not apply to the coherences $\rho_{n\neq m}$ which in general are
complex numbers. For a single qubit driven by one or several heat reservoirs
the coherences do not couple to the populations. The populations settle to their
steady state values and the coherences die out. For a composite system  the situation 
is different. Here coherences can in principle couple to populations and influence the 
steady state. This is for example the case in \cite{Li14,Zhang14,Wichterich07}.
Here we address this issue in the case of driven coupled qubits,

It is well-established that the proper master equation to be used in dealing with open
quantum systems is the Lindblad form \cite{Breuer06,Manzano20,Lindblad76,Chrus17}
which in the Markov limit ensures that the 
density matrix evolves in time according to a quantum dynamical semigroup defining 
a so-called quantum channel \cite{Wilde17}. The Lindblad master equation thus ensures that
the populations stay positive and that the trace condition $\sum_n\rho_{nn}(t)=0$
(conservation of probability) is satisfied. The Redfield master equation \cite{Redfield65} 
introduced 
prior to the Lindblad form preserves the trace but does not represent a quantum channel 
and can lead to unphysical negative populations; note that the trace condition does not
in itself ensure positive populations.

In recent work we applied a field theoretical condensed matter approach to open
quantum system 
\cite{Fogedby22}. 
Choosing a Caldeira-Leggett multi oscillator heat bath
\cite{Caldeira83a,Caldeira83b}
and
invoking Wick's theorem we derived a Dyson equation for the transmission 
operator $T$ propagating the density operator according to $\rho(t)=T(t,0)\rho(0)$.
The Dyson equation incorporating secular effects has the schematic form 
$T=T_0+T_0KT$, where the kernel $K$ can be determined by a an expansion in
powers of the system-bath coupling in terms of Feynman diagrams. Finally, 
applying
a so-called  quasi particle approximation, well-known in condensed matter
many body theory, we derived a master equation in the Markov approximation.
In the field theoretical approach the Dyson equation replaces some of the
physical assumptions made in the standard derivation of a master equation.
In the Born approximation the field theoretical method basically corresponds
to the Redfield equation \cite{Redfield65,Breuer06} in an energy basis.
Whereas the trace condition is automatically
satisfied, the field theoretical approach does not define a quantum channel,
unless a further selection rule is imposed. The selection rule corresponds to
the rotating wave approximation used in turning the Redfield equation into
the Lindblad equation in the standard derivation  \cite{Breuer06}. In more detail, 
the system 
operator $S^\alpha$  in an energy basis,
$S^\alpha_{nn'}$, appears in a bilinear combination $S^\alpha_{nn'}S^\beta_{pp'}$ 
in both the Redfield and Lindblad cases. As discussed in \cite{Fogedby22}, 
the selection rule enters as a constraint on the energy transitions according to
$S^\alpha_{nn'}S^\beta_{pp'}\delta(E_{nn'}+E_{pp'})$, where $E_{nn'}=E_n-E_{n'}$ is
the energy associated with the transition $n\to n'$. In a physical interpretation the
selection rule implies that the transition energies in the bilinear terms have
to match. It is also clear that whether the selection rule imposes a constraint
depends on the matrix elements $S^\alpha_{nn'}$. In cases where the selection
rule is trivially satisfied the Redfield and Lindblad approaches yield identical
density matrices. 

In the present paper we find that a Redfield approach to the case of coupled qubits
does yield coherence contribution to the steady state and, moreover, gives rise
to unphysical negative populations. In a proper Lindblad approach, corresponding
to a quantum channel, the populations are positive and the heat currents
are in accordance with the second law of thermodynamics.
We note that the present general analysis included for the sake of completion
has a strong overlap with the analysis in \cite{Soret22}.
The paper is organised as follows. In Sec. \ref{general} we discuss i) the master
equation, ii) the heat currents, and iii) the Redfield - Lindblad approaches and coherences.
In Sec. \ref{systems} we consider i) the single qubit case and ii) the coupled-qubit case.
In Sec. \ref{redfield} we show that a Redfield approach yields unphysical negative 
populations. In Sec. \ref{discussion} we give a brief discussion and summary.
Details regarding the Redfield approach is deferred to an Appendix \ref{appendix}
\section{\label{general}General Analysis}
Here we set up the general scheme for the evaluation of the density matrix and heat 
currents for an open system driven by two heat reservoirs.
\subsection{Master equation}
The density operator for an open quantum system $S$ characterised by the Hamiltonian 
$H$ interacting with a single or several heat reservoirs is in the Markov approximation 
governed by a master equation of the form
\cite{Breuer06,Rivas12,Fogedby22}
\begin{eqnarray}
\frac{d\rho(t)}{dt}=-i[H,\rho(t)]+L\rho(t).
\label{Master}
\end{eqnarray}
The von Neumann term $-i[H,\rho(t)]$ yields the  reversible unitary time evolution
\cite{vonNeumann27}.
The second term driven by the super operator $L$ characterises the coupling to
the environment. This term has the structure $L=K+i\Delta$, where $\Delta$ is a 
Lamb type correction to the energy levels. The dissipative kernel $K$ characterises 
the relaxation of the system. In the following we assume that the shift $\Delta$ is 
incorporated in $H$ and focus on the dissipative kernel $K$. 

In an energy basis, which we shall use in the following, we have
$H|n\rangle=E_n|n\rangle$, where $|n\rangle$ denotes the wavefunction and $E_n$
the associated energy. Expanding the von Neumann 
term the master equation in (\ref{Master}) takes the form
\begin{eqnarray}
\frac{d\rho(t)_{pp'}}{dt}=-iE_{pp'}\rho(t)_{pp'}+\sum_{qq'}K_{pp',qq'}\rho(t)_{qq'},
\label{EMaster}
\end{eqnarray}
where the energy shift is given by $E_{pp'}=E_p-E_{p'}$, the density matrix
by $\rho(t)_{pp'}=\langle p|\rho(t)|p'\rangle$, and the kernel super matrix
by $K_{pp',qq'}=\langle p|\langle q'|K|q\rangle |p'\rangle$. We are, moreover,
in the following assuming that the Markov approximation, i.e., the separation of
time scales, applies and that the kernel $K$ is time independent.

Consequently, the steady state density matrix denoted by $\rho_{pp'}$ is 
determined by a set of linear equations
\begin{eqnarray}
-iE_{pp'}\rho_{pp'}+\sum_{qq'}K_{pp',qq'}\rho_{qq'}=0;
\label{Steady-State}
\end{eqnarray}
note that the trace condition $\sum_n\rho_{nn}(t)=1$ implies $\sum_nd\rho_{nn}(t)/dt=0$ 
and thus
\begin{eqnarray}
\sum_{p}K_{pp,qq'}=0.
\label{Trace}
\end{eqnarray}
Separating (\ref{Steady-State}) with respect to 
populations and coherences, respectively,  we obtain the
coupled linear equations
\begin{eqnarray}
&&\sum_{q} K_{pp,qq}\rho_{qq}+\sum_{q\neq q'} K_{pp,qq'}\rho_{qq'}=0,
\label{Pop}
\\
&&-iE_{pp'}\rho_{pp'}+\sum_{q\neq q'}K_{pp',qq'}\rho_{qq'}+
\sum_{q}K_{pp',qq}\rho_{qq}=0, \text{for}~  p\neq p',
\label{Coh}
\end{eqnarray}
where in (\ref{Pop}) the populations $\rho_{qq}$ are coupled to the coherences
$\rho_{q\neq q'}$ driving the populations in (\ref{Coh}). Thus, in order for coherences
to influence the population we must have $K_{pp,q\neq q'}\neq 0$.
\subsection{Currents}
 The mean energy of the system is given by $E(t)=\text{Tr}(H\rho(t))$.
 For the energy or heat flux we thus have $q(t)=\text{Tr}(H d\rho(t)/dt)$
 or inserting (\ref{Steady-State}) 
\begin{eqnarray}
q=\sum_{n,qq'}E_nK_{nn,qq'}\rho_{qq'}.
\label{Cur}
\end{eqnarray}
For a single heat reservoir the system equilibrates and it it follows directly 
from (\ref{Steady-State}) that the heat current $q$ vanishes.
However, in the presence of several heat reservoirs maintained at different 
temperatures  a steady state heat current will be generated. 

In the case of two uncorrelated heat reservoirs  $A$ and $B$ the kernel $K$ 
in (\ref{Steady-State}) is composed of two parts associated with each reservoir 
and we have
\begin{eqnarray}
K_{pp',qq'}=K_{pp',qq'}^A+K_{pp',qq'}^B;
\label{Ker}
\end{eqnarray}
note that in the case where the heat bath are correlated there will an additional
term $K_{pp',qq'}^{A,B}$ depending on both heat baths.

As a result the  total current is determined by 
\begin{eqnarray}
q=\sum_{n,qq'}E_n(K_{nn,qq'}^A+K_{nn,qq'}^B)\rho_{qq'}.
\label{Heat}
\end{eqnarray}
The vanishing total current has components associated with each bath, i.e.,
\begin{eqnarray}
&&q=q^A+q^B=0,
\label{TotalHeat}
\\
&&q^{A}=\sum_{n,qq'}E_nK_{nn,qq'}^{A}\rho_{qq'},
\label{Acur}
\\
&&q^{B}=\sum_{n,qq'}E_nK_{nn,qq'}^{B}\rho_{qq'}.
\label{Bcur}
\end{eqnarray}
We infer that the heat flux from the reservoir $A$ to the system $S$ given by $q^A$
equals the heat flux $-q^B=q^A$ flowing from the system $S$ into the reservoir $B$,
expressing energy conservation, i.e. the first law of thermodynamics. Regarding
the second law of thermodynamics the issue is to show that a positive heat current
flows from the hot heat reservoir to the cold heat reservoir.
From the above it follows that in order to evaluate the heat currents for a given 
open quantum system driven by two reservoirs we need three components:  
The energy spectrum of the system $E_n$, the kernel elements $K^{A,B}_{nn,qq'}$,  
and the density operator $\rho_{qq'}$. 
\subsection{Redfield - Lindblad - Coherences}
Then issue of coherences in the steady state and currents is subtle 
\cite{Li14,Zhang14,Wichterich07}.
In recent work 
\cite{Fogedby22} mentioned in the introduction we presented a detailed discussion 
of the application of field theoretical methods to open quantum systems. The field 
theoretical approach is based on a systematic expansion in terms of diagrams and 
applies a condensed matter quasi-particle approximation in order to implement the 
Markov limit. The field theoretical method is the basis for the present analysis.

Assuming a system - bath coupling of the form $\sum_\alpha S^\alpha B^{\alpha}$, 
where $S^\alpha$ are the system operators and  $ B^\alpha$ the bath operators, 
the dissipative kernel $K_{pp',qq'}$ in the Born approximation and in an energy basis 
is given by
\begin{eqnarray}
K^\text{Red}_{pp',qq'}=
&&-\delta_{p'q'}\frac{1}{2}\sum_{\alpha\beta,l}
S^\alpha_{pl}S^\beta_{lq}
D^{\alpha\beta}(E_{pl})
-\delta_{pq}\frac{1}{2}\sum_{\alpha\beta,l}
S^\alpha_{q'l}S^\beta_{lp'}
D^{\alpha\beta}(E_{p'l})
\nonumber
\\
&&+\frac{1}{2}\sum_{\alpha\beta}
S^\beta_{pq}S^\alpha_{q'p'}(
D^{\alpha\beta}(E_{q'p'})+D^{\alpha\beta}(E_{qp})).
\label{Field-Kernel-Redfield}
\end{eqnarray}
Here $D^{\alpha\beta}(\omega)$ is the bath correlation function given by
\begin{eqnarray}
&&D^{\alpha\beta}(\omega)=\int dt \exp(i\omega(t-t'))
\text{Tr}_B[\rho_B B^\alpha(t)B^\beta(t')],
\label{Corr-Function}
\\
&&\rho_B=\frac{\exp(-H_B/T)}{\text{Tr}_B[\exp(-H_B/T )]},
\label{Den-Bath}
\end{eqnarray}
where $H_B=\sum_k\omega_kb_k^\dagger b_k$ is the bath Hamiltonian,
$b_k$ the Bose operator associated with the wavenumber $k$, 
and $T$ the associated temperature.

By inspection of (\ref{Field-Kernel-Redfield}) it is easily seen that the trace condition in 
(\ref{Trace}) is satisfied. However, the field theoretical approach which
in the Born approximation is equivalent to the Redfield equation \cite{Redfield65}
does not represent the generator of a quantum dynamical semigroup and thus 
does not guarantee that the populations are positive. In order to implement the
semigroup properties and thus guarantee positive populations one must impose
a further secular rotating wave approximation (RWA) yielding the Lindblad equation
\cite{Breuer06}.

In operator form the Lindblad master equation has the general form 
\begin{eqnarray}
\frac{d}{dt}\rho(t)=-i[H,\rho(t)]+\sum_{\alpha\beta,k}\gamma_k^{\alpha\beta}
\Big[S^\beta_k\rho(t)S_k^{\alpha\dagger}-
\frac{1}{2}\{S_k^{\alpha\dagger} S_k^\beta,\rho(t)\}\Big].
\label{Lindblad}
\end{eqnarray}
However, as shown in \cite{Fogedby22} the introduction of the RWA amounts
to introducing delta function constraints in the kernel  (\ref{Field-Kernel-Redfield})
yielding the expression
\begin{eqnarray}
K^\text{Lind}_{pp',qq'}=
&&-\frac{\delta_{p'q'}}{2}\sum_{\alpha\beta,l}
S^\alpha_{pl}S^\beta_{lq}\delta(E_{pl}+E_{lq})
D^{\alpha\beta}(E_{pl})
-\frac{\delta_{pq}}{2}\sum_{\alpha\beta,l}
S^\alpha_{q'l}S^\beta_{lp'}\delta(E_{q'l}+E_{lp'})
D^{\alpha\beta}(E_{p'l})
\nonumber
\\ &&+\frac{1}{2}\sum_{\alpha\beta}
S^\beta_{pq}S^\alpha_{q'p'}\delta(E_{pq}+E_{q'p'})
(D^{\alpha\beta}(E_{q'p'})+D^{\alpha\beta}(E_{qp})).
\label{Field-Kernel-Lindblad}
\end{eqnarray}
In the interaction representation the system operator 
$S^\alpha_{pq}(t)\propto\exp(-iE_{pq}t)$.
Consequently, the product 
$S^\alpha_{pq}(t)S^\beta_{p'q'}(t)\propto\exp(-i(E_{pq}+E_{p'q'})t)$.
For $E_{pq}+E_{p'q'}\neq 0$ this term will oscillate and is discarded in the RWA. 
As a result, whether the delta function constraint eliminates terms in the super operator
$K_{pp',qq'}$ depends on the matrix elements $S^\alpha_{pq}$ and thus on the specific
model under consideration. 

We stress that a proper analysis satisfying the correct physics requires the Lindblad
equation. Whether the Lindblad equation then generates coherences in the steady state
will, as mentioned above, depend on the specific model 
\cite{Li14,Zhang14,Wichterich07}. In the case of a single driven 
qubit it follows immediately that coherences are absent in the steady state. In the coupled
qubit case we find that a Redfield treatment according to  (\ref{Field-Kernel-Redfield})
yield coherences in the steady state and unphysical negative populations. On the other hand, 
for coupled qubits a 
Lindblad treatment according to (\ref{Field-Kernel-Lindblad}) yields vanishing coherences,
positive populations and simple expressions for the currents.
\section{\label{systems}Qubit systems driven by two heat reservoirs}
In this section we turn to the case of qubit systems driven by two heat reservoirs generating
a non equilibrium heat current. As an example and illustration we first consider the case of a 
driven single qubit. The second case, constituting the main issue in the paper, is a coupled
qubit system.
\subsection{\label{qubit}Single qubit}
\begin{figure}
\centering
\includegraphics[width=1.0\hsize]{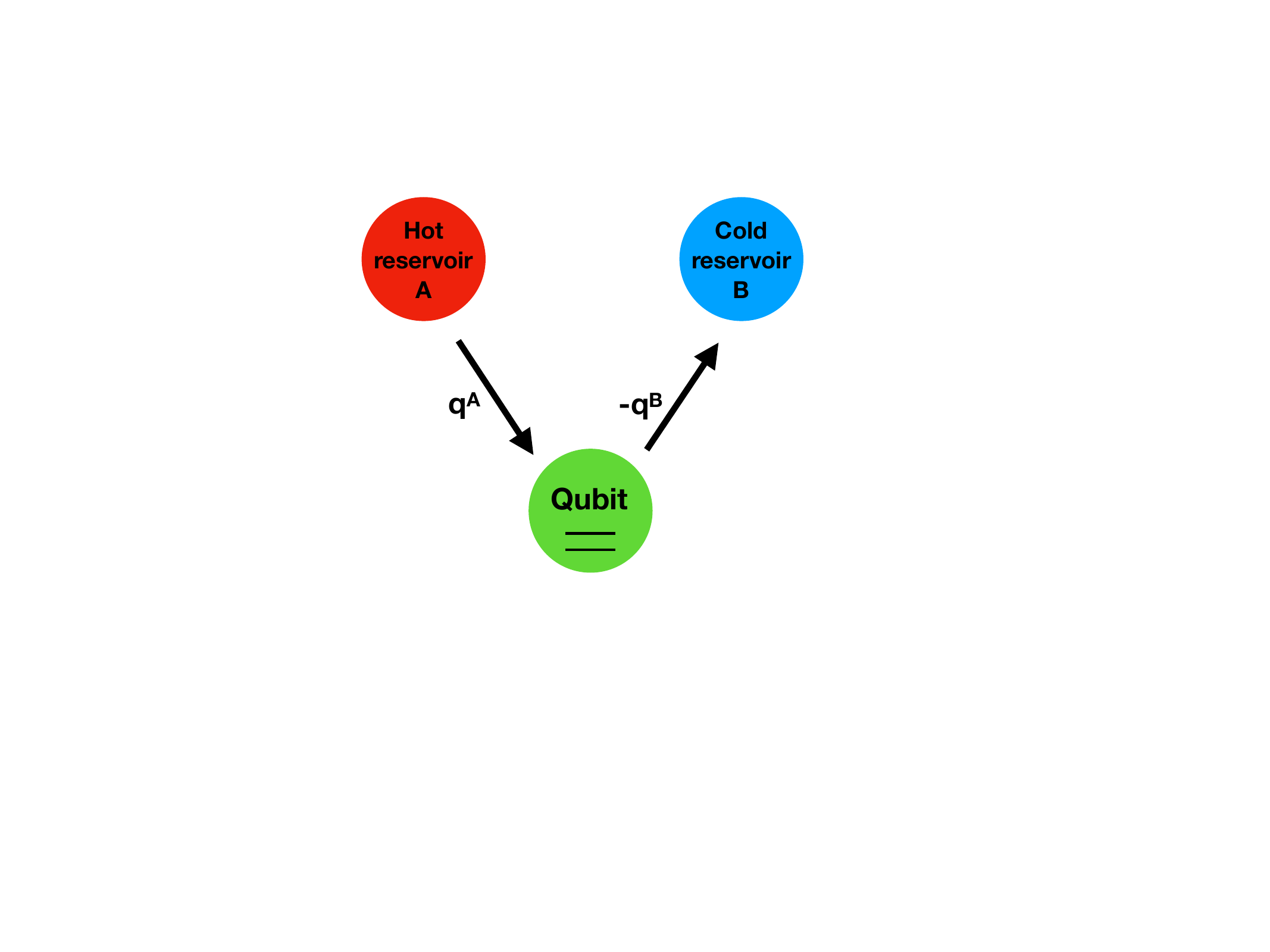}
\caption{
We depict the drive of a single qubit by a hot heat reservoir $A$ and a
cold heat reservoir $B$.}
\label{fig1}
\end{figure}

Here we consider the case of a single qubit driven by two heat reservoirs $A$ and $B$
maintained at temperatures $T_A$ and $T_B$, respectively. In this case the energy or 
heat flows from the hot reservoir to the cold reservoir via the qubit; the configuration is 
depicted in Fig.~\ref{fig1}.
\subsubsection{Model}
In a Pauli matrix basis 
\cite{Zinn-Justin89} 
the Hamiltonian and the coupling to the reservoirs are given by
\begin{eqnarray}
&&H=\frac{\omega_0}{2}\sigma^z
\label{HQubit}
\\
&&H_{A}=\sigma^+A+\sigma^-A^\dagger,
\label{HA}
\\
&&H_{B}=\sigma^+B+\sigma^-B^\dagger,
\label{HB}
\\
&&A=\sum_k\lambda_k^A a_k,
\label{A}
\\
&&B=\sum_k\lambda_k^B b_k,
\label{B}
\end{eqnarray}
where the levels $|+\rangle$ and $|-\rangle$ have energies $E_+=\omega_0/2$
and $E_-=-\omega_0/2$, corresponding to the energy splitting $\omega_0$. 
The qubit is driven by two heat reservoirs in the rotating wave approximation (RWA). 
The bath operators $A$ and $B$ sample the bath modes with
wavenumber $k$, frequencies $\omega_k^{A,B}$, and strength $\lambda_k^{A,B}$;
$a_k$ and $b_k$ are Bose operators. 

Referring to Sec. \ref{general} we choose the assignment  
$S^\alpha=(S^1,S^2)=(\sigma^+,\sigma^-)$, $A^\alpha=(A^1,A^2)=(A,A^\dagger)$,
and $B^\alpha=(B^1,B^2)=(B,B^\dagger)$. For the bath correlations we find 
\begin{eqnarray}
&&D^{12}_{A,B}(\omega)=g_{A,B}(\omega)(1+n_{A,B}(\omega)),~~\omega>0,
\label{Corr1}
\\
&&D^{21}_{A,B}(\omega)=g_{A,B}(-\omega)n_{A,B}(-\omega),~~~~~\omega<0,
\label{Corr2}
\end{eqnarray}
with spectral functions and Planck distributions
\begin{eqnarray}
&&g_{A,B}(\omega)=2\pi\sum_k(\lambda_k^{A,B})^2\delta(\omega-\omega^{A,B}_k),
\label{Spec}
\\
&&n_{A,B}(\omega)=\frac{1}{\exp(\omega/T_{A,B})-1}.
\label{Planck}
\end{eqnarray}
Inserting the nonvanishing matrix elements $\langle +|\sigma^+|-\rangle=
\langle -|\sigma^-|+\rangle=1$ with associated energy shifts $E_{+-}=-E_{-+}=\omega_0$
we obtain from (\ref{Field-Kernel-Lindblad}) with the abbreviations 
$g_{A,B}=g_{A,B}(\omega_0)$ and $n_{A,B}=n_{A,B}(\omega_0)$ the 
dissipative kernel elements
\begin{eqnarray}
&&K^{A,B}_{++,++}=-g_{A,B}(1+n_{A,B}),
\label{kernel1}
\\
&&K^{A,B}_{--,++}=+g_{A,B}(1+n_{A,B}),
\label{kernel2}
\\
&&K^{A,B}_{++,--}=+g_{A,B}n_{A,B},
\label{kernel3}
\\
&&K^{A,B}_{--,--}=-g_{A,B}n_{A,B},
\label{kernel4}
\\
&&K^{A,B}_{+-,+-}=-\frac{g_{A,B}}{2}(1+2n_{A,B}),
\label{kernel5}
\\
&&K^{A,B}_{-+,-+}=-\frac{g_{A,B}}{2}(1+2n_{A,B});
\label{kernel6}
\end{eqnarray}
we note that the delta function constraints in (\ref{Field-Kernel-Lindblad}) are trivially
satisfied showing that the Redfield and Lindblad schemes yield identical kernel elements.
Note also that there is no coupling between populations and coherences, i.e.
$K_{nn,q\neq q'}=0$. 
\subsubsection{Density matrix}
Expressing the kernel $K$ in matrix form 
\begin{eqnarray}
K=
\left(\begin{array}{cccc}
K_{++,++} & K_{++,--} & 0 & 0
\\
K_{--,++} & K_{--,--} & 0 & 0
 \\
 0 & 0 & K_{+-,+-} & 0
  \\
  0 & 0 &0 & K_{-+,-+}
 \end{array}\right),
 \label{Matrix-1}
\end{eqnarray}
we have (suppressing indices)
\begin{eqnarray}
K=
\left(\begin{array}{cccc}
-g(1+n) & gn & 0 & 0
\\
g(1+n) & -gn & 0 & 0
 \\
 0 & 0 & -(g/2)(1+2n) & 0
  \\
  0 & 0 &0 & -(g/2)(1+2n)
 \end{array}\right),
 \label{Matrix-2}
\end{eqnarray}
forming a "population" block $P$ and a "coherence" block $C$. For the determinant
of $K$ we thus have $\det K=\det P\det C$, where $\det P=0$ and 
$\det C=(g(1+2n))^2/4\neq 0$. Consequently, the populations $\rho_{nn}$ determined
by the linear equations $P\rho=0$, requiring a vanishing determinant, yield finite
populations, whereas the coherences determined by $C\rho=0$ vanish in the steady
state since the determinant is non vanishing. Applying the scheme in Sec. \ref{general}
we readily obtain the populations or diagonal density matrix elements
\begin{eqnarray}
&&\rho_{++}=
\frac{g_An_A+
g_Bn_B}
{g_A(1+2n_A)+g_B(1+2n_B)},
\label{rhop}
\\
&&\rho_{--}=
\frac{g_A(1+n_A)+
g_B(1+n_B)}
{g_A(1+2n_A)+g_B(1+2n_B)}.
\label{rhom}
\end{eqnarray}
A few comments regarding the density matrix.
Introducing the ratio $r= \rho_{++}/\rho_{--}$ we have from (\ref{rhop}) and (\ref{rhom})
\begin{eqnarray}
r=\frac{g_An_A+g_Bn_B}{g_A(1+n_A)+g_B(1+n_B)}.
\label{ratio}
\end{eqnarray}
In general the ratio $r$ depends on both the spectral densities $g_{A,B}$,
i.e., the structure of the reservoirs, and the temperatures $T_{A,B}$. Turning 
off for example reservoir $B$ by setting $g_B=0$ we obtain the well-studied 
case of a single qubit driven by a single reservoir \cite{Breuer06}. The 
ratio $r$ does not depend on the spectral strength and we obtain 
$r=\exp(-\omega_0/T_A)$, i.e., the Boltzmann distribution. For identical reservoirs
with the same spectral densities, i.e., $g_A=g_B=g$, we obtain 
$r=(n_A+n_B)/(2+n_A+n_B)$. At high temperatures $r\to 1-\omega_0/T$, where 
$T=(T_A+T_B)/2$ is the mean temperature of the combined reservoirs and 
the energy levels become equally populated. In the low temperature limit 
$r\to (\exp(-\omega_0/T_A)+ \exp(-\omega_0/T_B))/2$. Introducing the temperature 
bias $\Delta=T_A-T_B$ we  have $r\to\exp(-\omega_0/T)\cosh(2\omega_0/\Delta)$, 
i.e., a correction to the Boltzmann factor. At high temperature, $T\gg\omega_0$, 
we have $\rho_{+}\to 1/2$ and $\rho_{-}\to 1/2$, i.e., equal populations; at low 
temperature, $T\ll\omega_0$, we have $\rho_{+}\to 0$ and $\rho_{-}\to 1$, i.e.,
population of the lowest level (ground state). 
\subsubsection{Currents}
Noting that $E_+=\omega_0/2$ and $E_-=-\omega_0/2$ the scheme in Sec. \ref{general}
likewise yields expressions for the currents. The  steady state heat currents from reservoir 
$A$ and from reservoir $B$ to the qubit are thus given by
\begin{eqnarray}
&&q^A=
\frac{g_Ag_B\omega_0}
{g_A(1+2n_A)+
g_B(1+2n_B)}[n_A-n_B],
\label{singlequbitA}
\\
&&q^B=
\frac{g_Ag_B\omega_0}
{g_A(1+2n_A)+
g_B(1+2n_B)}[n_B-n_A].
\label{singlequbitB}
\end{eqnarray}
First we note that by construction $q^A+q^B=0$ showing that energy is conserved, i.e.,
the first law of thermodynamics. Secondly, since $n_A>n_B$ for $T_A>T_B$ energy flows
from the hot heat reservoir with temperature $T_A$ to the cold heat reservoir maintained at
temperature $T_B$, demonstrating that the second law of thermodynamics holds.

Regarding the currents given by (\ref{singlequbitA}) and  (\ref{singlequbitB}) we note
that they depend on both spectral representations $g^A$ and $g^B$. Removing for 
example reservoir $B$ by setting $g^B=0$ the currents vanish and the qubit coupled 
only to reservoir $A$ equilibrates with temperature $T_A$. In the high temperature limit for 
$T_{A,B}\gg\omega_0$ we obtain expanding (\ref{singlequbitA}) and (\ref{singlequbitB})
the classical results
\begin{eqnarray}
&&q^A\sim\frac{1}{2}\frac{g_Ag_B\omega_0}{g_AT_A+g_BT_B}[T_A-T_B],
\label{ahigh}
\\
&&q^B\sim\frac{1}{2}\frac{g_Ag_B\omega_0}{g_AT_A+g_BT_B}[T_B-T_A],
\label{bhigh}
\end{eqnarray}
at low temperatures for $T_{A,B}\ll\omega_0$ the vanishing currents
\begin{eqnarray}
&&q^A\sim\frac{g_Ag_B\omega_0}{g_A+g_B}\Big[e^{-\omega_0/T_A}-e^{-\omega_0/T_B}\Big],
\label{alow}
\\
&&q^B\sim\frac{g_Ag_B\omega_0}{g_A+g_B}\Big[e^{-\omega_0/T_B}-e^{-\omega_0/T_A}\Big].
\label{blow}
\end{eqnarray}
%
\subsection{\label{qubits} Coupled qubits}
\begin{figure}
\centering
\includegraphics[width=1.0\hsize]{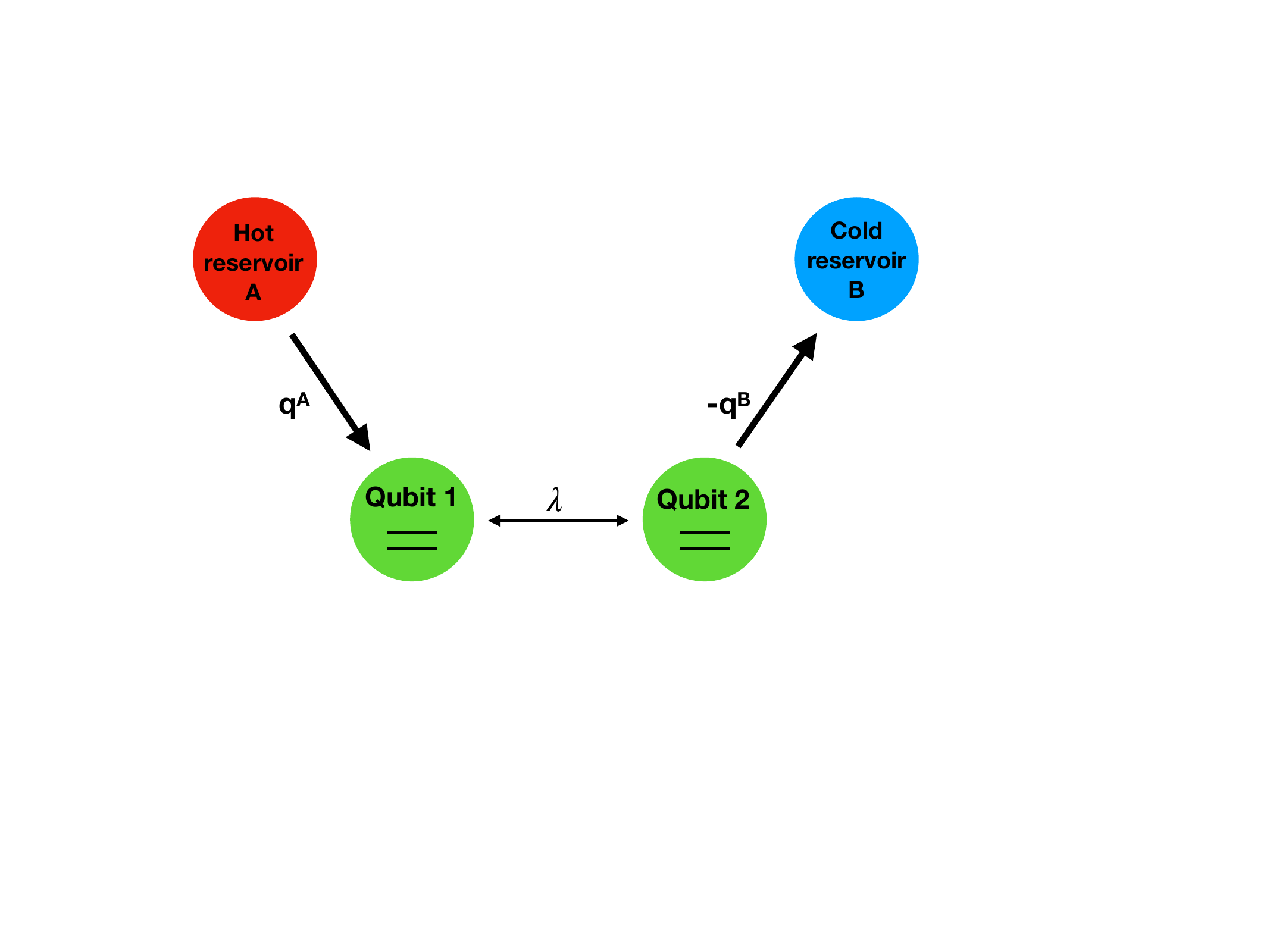}
\caption
{
We depict the drive of a composite two-qubit system by a hot heat reservoir $A$ 
coupled to qubit 1 and a cold heat reservoir $B$ coupled to qubit 2.
}
\label{fig2}
\end{figure}

This is the main section in the paper. We discuss in detail the case of two coupled qubits 
driven by two reservoirs
as a model of a simple composite system. Heat reservoir $A$ drives qubit 1 and heat 
reservoir $B$ drives qubit 2. In this case heat flows between the reservoirs and the qubits 
transmitted across the system by the qubit coupling; the configuration is depicted 
in Fig.~\ref{fig2}. 
\subsubsection{Model}
In a Pauli matrix basis  
\cite{Zinn-Justin89} 
the Hamiltonian and coupling to the reservoirs are given by
\begin{eqnarray}
&&H=\frac{\omega_1}{2}\sigma^z_1+\frac{\omega_2}{2}\sigma^z_2+
\lambda(\sigma^+_1\sigma^-_2+ \sigma^-_1\sigma^+_2).
\label{HQubit2}
\\
&&H_{A}=\sigma_1^+A+\sigma_1^-A^\dagger,
\label{HA2}
\\
&&H_{B}=\sigma_2^+B+\sigma_2^-B^\dagger,
\label{HB2}
\end{eqnarray}
where $\omega_1$ and $\omega_2$ are the energy splittings of the unperturbed qubits
and $\lambda$ the coupling strength of the flip-flop interaction, i.e., the dipole coupling
in the RWA. For the composite system we have the 
unperturbed states $|--\rangle$, $|-+\rangle$, $|+-\rangle$ and $|++\rangle$  
with energies $-\omega_m$, $-\Delta\omega$, $\Delta\omega$, and $\omega_m$,
respectively, where
\begin{eqnarray}
&&\omega_m=\frac{\omega_1+\omega_2}{2},
\label{om}
\\
&&\Delta\omega=\frac{\omega_1-\omega_2}{2}.
\label{del}
\end{eqnarray}
%

Diagonalising $H$ yields the four states 
\begin{eqnarray}
&&|1\rangle=|--\rangle,
\label{st1}
\\
&&|2\rangle=-\beta|+-\rangle+\alpha|-+\rangle,
\label{st2}
\\
&&|3\rangle=\alpha|+-\rangle+\beta|-+\rangle,
\label{st3}
\\
&&|4\rangle=|++\rangle,
\label{st4}
\end{eqnarray}
with energies
\begin{eqnarray}
&&E_1=-\omega_m, 
\label{e1}
\\
&&E_2=-\sqrt{(\Delta\omega)^2+\lambda^2},
\label{e2}
\\
&&E_3= \sqrt{(\Delta\omega)^2+\lambda^2},
\label{e3}
\\
&&E_4=\omega_m.
\label{e4}
\end{eqnarray}
The expansion coefficients and relevant energy shifts 
$\omega_+=E_{42}=E_{31}$ and $\omega_-=E_{43}=E_{21}$
are given by
\begin{eqnarray}
&&\alpha=\cos\theta/2, 
\label{alpha}
\\
&&\beta=\sin\theta/2, 
\label{beta}
\\
&&\tan\theta=2\lambda/(\omega_1-\omega_2), ~~ 0<\theta<\pi/2,
\label{tan}
\\
&&\omega_+=\omega_m + \sqrt{(\Delta\omega)^2+\lambda^2},
\label{omp}
\\
&&\omega_-=\omega_m - \sqrt{(\Delta\omega)^2+\lambda^2}.
\label{omm}
\end{eqnarray}
For vanishing coupling between the qubits, $\lambda\to 0$, we have 
$\alpha\to 1$, $\beta\to 0$, $\omega_+\to \omega_1$,
and $\omega_-\to \omega_2$;
moreover, assuming $\omega_\pm>0$ we require $\lambda<\sqrt{\omega_1\omega_2}$.
\subsubsection{Kernel}
In order to evaluate the dissipative kernel according to the scheme in Sec. \ref{general}
we require the matrix elements of the system operators $\sigma_{1,2}^\pm$ in the energy
basis $|n\rangle$, $n=1\cdots 4$ of the system Hamiltonian $H$ in (\ref{HQubit2}).
With the notation $(\sigma_{1,2}^\pm)_{pq}=\langle p|\sigma_{1,2}^\pm |q\rangle$ we find 
the non-vanishing matrix elements and associated energy transitions
\begin{eqnarray}
\left(\begin{array}{cc}
(\sigma^+_1)_{31} & E_{31} 
\\
(\sigma^+_1)_{42} & E_{42}
\\
(\sigma^+_1)_{43} & E_{43}
\\
(\sigma^+_1)_{21} & E_{21}\end{array}\right)
=
\left(\begin{array}{cc}
\alpha & \omega_+
 \\
\alpha &  \omega_+
\\
\beta &  \omega_-
 \\
 -\beta &  \omega_- 
 \end{array}\right),
 \label{M1}
\end{eqnarray}
and
\begin{eqnarray}
\left(\begin{array}{cc}
(\sigma^+_2)_{21} & E_{21} 
\\
(\sigma^+_2)_{43} & E_{43}
\\
(\sigma^+_2)_{31} & E_{31}
\\
(\sigma^+_2)_{42} & E_{42}
\end{array}\right)
=
\left(\begin{array}{cc}
\alpha &  \omega_-
 \\
 \alpha &  \omega_-
\\
\beta &  \omega_+
 \\
 -\beta &  \omega_+
 \end{array}\right).
 \label{M2}
\end{eqnarray}
We note that reservoir $A$ drives the $\sigma_1$ transitions, whereas reservoir $B$
drives the  $\sigma_2$ transitions. Moreover, the system operators only appear in 
the combinations $\sigma_1^+\sigma_1^-$ and $\sigma_2^+\sigma_2^-$
in the kernel. We also note that the transitions fall in two groups. The transitions 
$1\leftrightarrow 3$ and $2\leftrightarrow 4$ are associated with $\omega_+$,
whereas the transitions $1\leftrightarrow 2$ and $3\leftrightarrow 4$ are associated
with $\omega_-$.

In order to identify nonvanishing kernel elements we inspect the expressions 
(\ref{Field-Kernel-Redfield}) in the Redfield case and (\ref{Field-Kernel-Lindblad})
in the Lindblad case. We introduce the positive parameters $a_n,b_n$ depending on the 
coupling strength, spectral densities, Planck distributions and temperatures according
to the scheme
\begin{eqnarray}
\left(\begin{array}{cc}
a_1 & b_1
\\
a_2 & b_2
\\
a_3 & b_3
\\
a_4 & b_4
\end{array}\right)
=
\left(\begin{array}{cc}
\alpha^2 g_A(\omega_+)(1+n_A(\omega_+)) & \beta^2g_B(\omega_+)(1+n_B(\omega_+))
\\
\beta^2g_A(\omega_-)(1+n_A(\omega_-)) & \alpha^2g_B(\omega_-)(1+n_B(\omega_-))
\\
\alpha^2g_A(\omega_+)n_A(\omega_+) & \beta^2g_B(\omega_+)n_B(\omega_+)
\\
\beta^2g_A(\omega_-)n_A(\omega_-) & \alpha^2g_B(\omega_-)n_B(\omega_-)
\end{array}\right).
\label{param1}
\end{eqnarray}
Using the assignment
\begin{eqnarray}
K=
\left(
\begin{array}{cccc}
K_{11,11}&K_{11,22}&K_{11,33}&K_{11,44}
\\
K_{22,11}&K_{22,22}&K_{22,33}&K_{22,44}
\\
K_{33,11}&K_{33,22}&K_{33,33}&K_{33,44}
\\
K_{44,11}&K_{44,22}&K_{44,33}&K_{44,44}
\nonumber
\end{array}
\right),
\label{K}
\end{eqnarray}
we obtain, since the delta function constraints are trivially satisfied, in both the 
Redfield and Lindblad cases the kernels $K^A$ and $K^B$:
\begin{eqnarray}
K^A=
\left(
\begin{array}{cccc}
-(a_3+a_4) &a_2&a_1 & 0
\\
a_4 &-(a_2+a_3) & 0 &a_1
\\
a_3 & 0 &-(a_1+a_4) & a_2
\\
0 &a_3&a_4&-(a_1+a_2)
\end{array}
\right),
\label{KA}
\end{eqnarray}
and
\begin{eqnarray}
K^B=
\left(
\begin{array}{cccc}
-(b_3+b_4) &b_2&b_1 & 0
\\
b_4 &-(b_2+b_3) & 0 &b_1
\\
b_3 & 0 &-(b_1+b_4) & b_2
\\
0 &b_3&b_4&-(b_1+b_2)
\end{array}
\right).
\label{KB}
\end{eqnarray}
In the Lindblad case, due to the delta function constraints, the above expressions
for $K^{A,B}$,
yielding populations and currents, exhaust the class of non vanishing kernel elements.

In the Redfield case, relaxing the delta function selection rules, we find non vanishing
expressions for kernel elements of the form  $K^{A,B}_{pp,q\neq q'}$,  
 $K^{A,B}_{p\neq p',qq}$, and  $K^{A,B}_{p\neq p',q\neq q'}$ indicating that the 
 populations couple to the coherences. The Redfield case will be discussed in
 the next section. However, since the Lindblad equation represent a proper quantum 
 channel we proceed with the evaluation of populations and currents.
\subsubsection{Density matrix}
Introducing the total kernel $K=K^A+K^B$, where $K^A$ and $K^B$ are given by 
(\ref{KA}) and (\ref{KB}), respectively, the populations $\rho_{nn}$ are determined
by the linear equations $\sum_q K_{pp,qq}\rho_{qq}=0$. By inspection we note 
incidentally that $\det K=0$ allowing for a solution. In terms of the parameters 
$s_n=a_n+b_n$, $n=1\cdots 4$, where
$a_n, b_n$ are given (\ref{param1}),
we find the normalised populations, i.e., $\sum_n\rho_{nn}=1$,
\begin{eqnarray}
\left(\begin{array}{c}
\rho_{11}\\\rho_{22}\\\rho_{33}\\\rho_{44}
\end{array}\right)
=\frac{1}{(s_1+s_3)(s_2+s_4)}
\left(\begin{array}{c}
s _1s_2 \\s_1s_4 \\s_3s_2 \\s_3s_4
\end{array}\right), ~~s_n=a_n+b_n.
\label{pop1}
\end{eqnarray}

In the high temperature limit $T_A, T_B\gg\omega_+, \omega_-$
we have $n_{A,B}(\omega_\pm)\to T_{A,B}/\omega_\pm$
and a simple calculation yields $\rho_{nn}\to 1/4$,
i.e., equal population of the four energy levels. Likewise, for 
$T_A, T_B\ll\omega_+, \omega_-$ we have $\rho_{11}\to 1$ and
$\rho_{22}, \rho_{33}, \rho_{44}\to 0$, i.e., occupation of the lowest energy level.

Finally, for $\lambda=0$ we have $\omega_+\to \omega_1$, $\omega_-\to \omega_2$,
$s_1=a_1,s_2=b_2,s_3=a_3,s_4=b_4$ and the populations take the form
\begin{eqnarray}
\left(\begin{array}{c}
\rho_{11}\\\rho_{22}\\\rho_{33}\\\rho_{44}
\end{array}\right)
=\frac{1}{(a_1+a_3)(b_2+b_4)}
\left(\begin{array}{c}
a _1b_2 \\a_1b_4 \\a_3b_2 \\a_3b_4
\end{array}\right).
\label{pop2}
\end{eqnarray}
By inspection we conclude that the qubits decouple and we obtain for the density operator
the direct product $\rho=\rho^A\otimes\rho^B$, where
\begin{eqnarray}
\rho^A=\frac{1}{g_A(\omega_1)(1+2n_ A(\omega_1))}
\left(\begin{array}{cc}g_A(\omega_1)(1+n_ A(\omega_1))& 0
 \\
 0 &g_A(\omega_1)n_ A(\omega_1)\end{array}\right),
 \label{Adirect}
\end{eqnarray}
and
\begin{eqnarray}
\rho^B=\frac{1}{g_B(\omega_2)(1+2n_ B(\omega_2))}
\left(\begin{array}{cc}g_B(\omega_2)(1+n_ B(\omega_2))& 0
 \\
 0 &g_B(\omega_2)n_ B(\omega_2)\end{array}\right),
 \label{Bdirect}
\end{eqnarray}
are the equilibrium density operators for qubit 1 coupled to reservoir $A$ and qubit 2
coupled to reservoir $B$, respectively. 

\begin{figure}
\centering
\includegraphics[width=0.9\hsize]{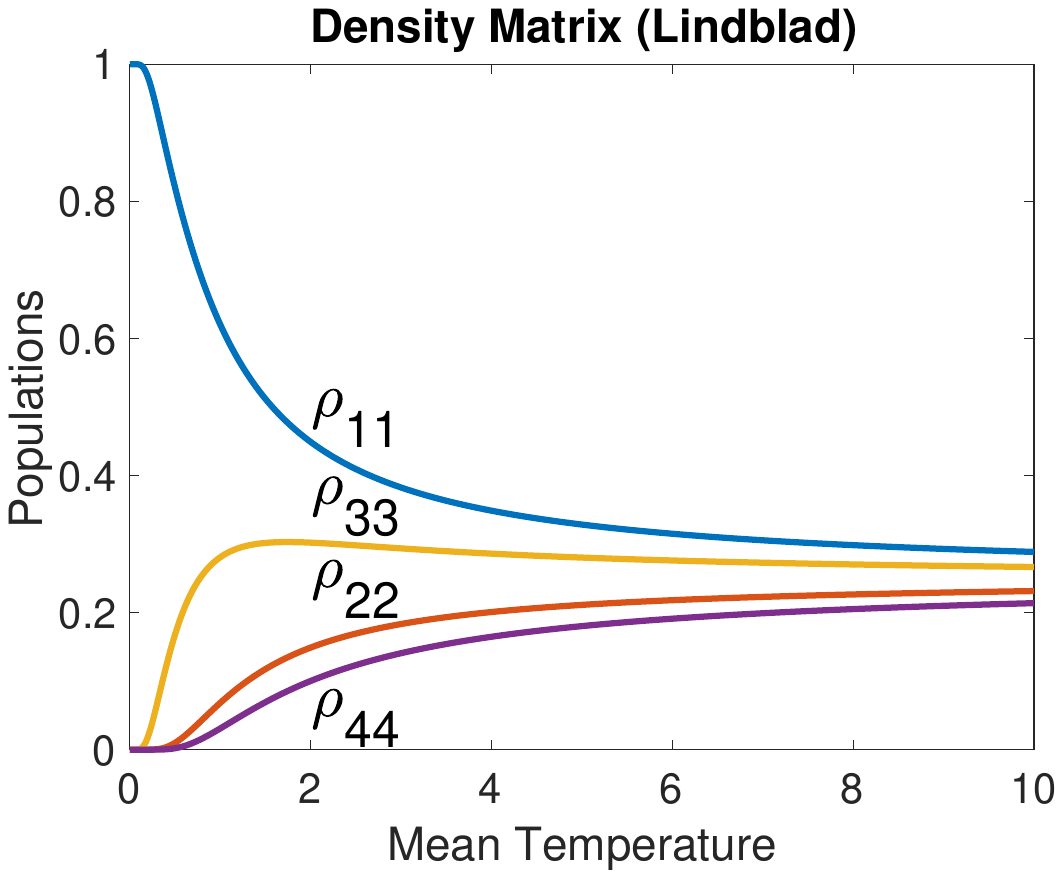}
\caption
{
We depict the populations $\rho_{11}$, $\rho_{22}$, $\rho_{33}$, $\rho_{44}$ 
in the Lindblad case as a function of the mean temperature $T$.
}
\label{fig3}
\end{figure}

In order to illustrate the Lindblad case we have in Fig.~\ref{fig3}  depicted 
the populations
$\rho_{nn}$ as a function of the mean temperature $T =(T_A+T_B)/2$ for $T_A=T_B$,
i.e. in the equilibrium case, with parameter choice $\omega_1=1.0$, $\omega_2=2.0$, 
$g_A(\omega_\pm)=g_B(\omega_\pm)=1.0$, and $\lambda=0.5$. For $T=0$ the 
system occupies the lowest level, i.e., $\rho_{11}=0$; at high temperatures the
populations converge, i.e., equal populations, $\rho_{nn}\to 1/4$.
\subsubsection{Currents}
Applying the scheme in Sec. \ref{general} the heat currents $q^A$ and $q^b$ are given by
$q^A=\sum_n E_nK^A_{nn,pp}\rho_{pp}$ and $q^B=\sum_n E_nK^B_{nn,pp}\rho_{pp}$
and we obtain the expressions
\begin{eqnarray}
&&q^A=\frac{a_3b_1-a_1b_3}
{s_1+s_3}\omega_+
+
\frac{a_4b_2-a_2b_4}
{s_2+s_4}\omega_-,
\label{Curr2A}
\\
&&q^B=\frac{a_1b_3-a_3b_1}
{s_1+s_3}\omega_+
+
\frac{a_2b_4-a_4b_2}
{s_2+s_4}\omega_-,
\label{Curr2B}
\end{eqnarray}
or by insertion of $a_n, b_n, s_n$
\begin{eqnarray}
&&q^A=q^A_++q^A_-,
\label{curA}
\\
&&q^A_+=\frac{(\alpha\beta)^2g_A(\omega_+)g_B(\omega_+)\omega_+}
{\alpha^2g_A(\omega_+)(1+2n_A(\omega_+))+
\beta^2g_B(\omega_+)(1+2n_B(\omega_+))}
[n_A(\omega_+)-n_B(\omega_+)],
\label{curAp}
\\
&&q^A_-=\frac{(\alpha\beta)^2g_A(\omega_-)g_B(\omega_-)\omega_-}
{\beta^2g_A(\omega_-)(1+2n_A(\omega_-))+
\alpha^2g_B(\omega_-)(1+2n_B(\omega_-))}
[n_A(\omega_-)-n_B(\omega_-)].
\label{curAm}
\end{eqnarray}
and
\begin{eqnarray}
&&q^B=q^B_++q^B_-,
\label{curB}
\\
&&q^B_+=\frac{(\alpha\beta)^2g_A(\omega_+)g_B(\omega_+)\omega_+}
{\alpha^2g_A(\omega_+)(1+2n_A(\omega_+))+
\beta^2g_B(\omega_+)(1+2n_B(\omega_+))}
[n_B(\omega_+)-n_A(\omega_+)],
\label{curBp}
\\
&&q^B_-=\frac{(\alpha\beta)^2g_A(\omega_-)g_B(\omega_-)\omega_-}
{\beta^2g_A(\omega_-)(1+2n_A(\omega_-))+\alpha^2g_B(\omega_-)(1+2n_B(\omega_-))}
[n_B(\omega_-)-n_A(\omega_-)].
\label{curBm}
\end{eqnarray}

By inspection we have $q_A+q_B=0$ expressing energy conservation, i.e., the first
law of thermodynamics. Moreover, and more importantly, since 
$n_A(\omega_\pm)>n_B(\omega_\pm)$ for $T_A>T_B$  the current expressions are in
accordance with the second law of thermodynamics, i.e., heat flows from the hot reservoir
to the cold reservoir; this result corroborates the analysis in \cite{Levy14} in the case of coupled
oscillators driven by two reservoirs.

\begin{figure}.
\centering
\includegraphics[width=0.9\hsize]{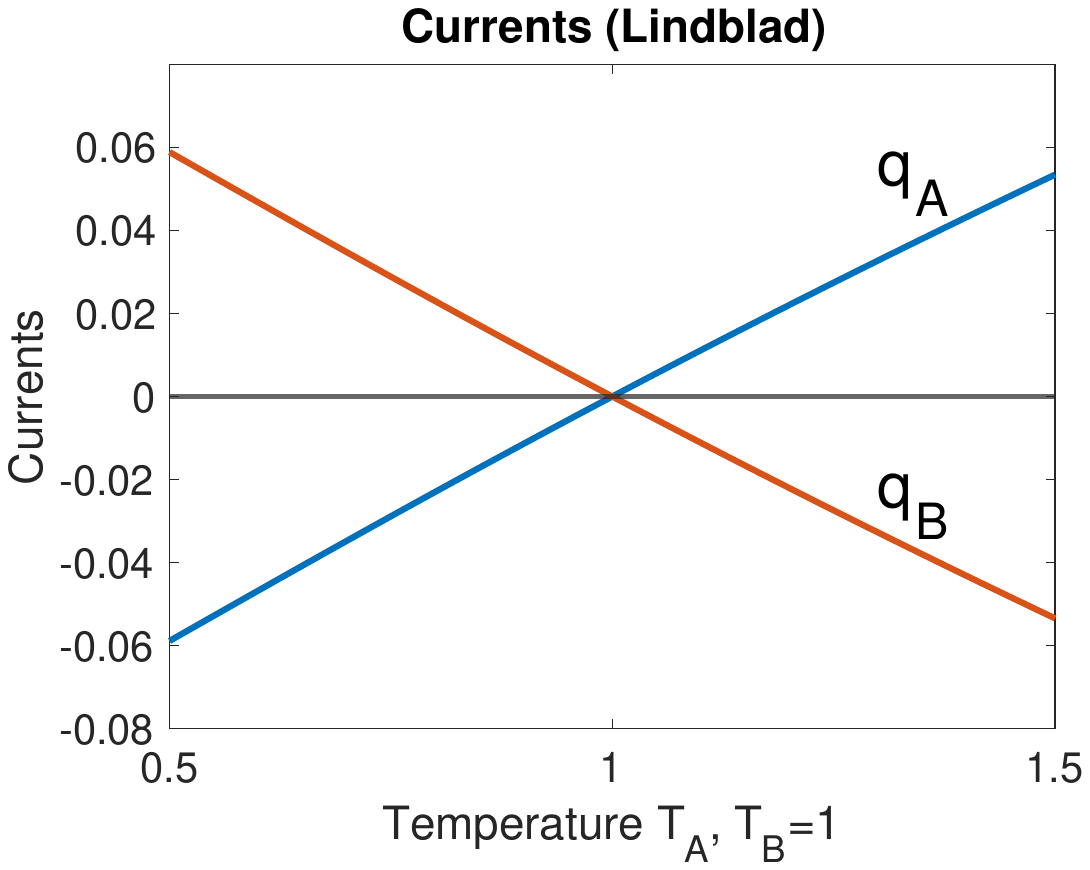}
\caption{We depict the heat currents $q_A$ and $q_B$ as functions
of $T_A$ for $T_B=1$.
}
\label{fig4}
\end{figure}

The heat currents $q^A$ and $q^B$ in (\ref{curA} - \ref{curBm}) fall in two parts associated
with the energy shifts $\omega_+$ and $\omega_-$. For vanishing coupling, $\lambda=0$,
we have $\alpha=1$ and  $\beta=1$, the qubits are uncoupled  and the currents vanish.
Likewise, quenching e.g. reservoir $B$ by setting $g_B=0$ we obtain a vanishing current.
In the high temperature limit for $T_A, T_B\gg\omega_+,\omega_-$ we have the
classical results
\begin{eqnarray}
&&q^A_+=\frac{1}{2}\frac{(\alpha\beta)^2g_A(\omega_+)g_B(\omega_+)\omega_+}
{\alpha^2g_A(\omega_+)T_A+\beta^2g_B(\omega_+)T_B}[T_A-T_B],
\label{curAAp}
\\
&&q^A_-=\frac{1}{2}\frac{(\alpha\beta)^2g_A(\omega_-)g_B(\omega_-)\omega_-}
{\beta^2g_A(\omega_-)T_A+
\alpha^2g_B(\omega_-)T_B}
[T_A-T_B].
\label{curAAm}
\end{eqnarray}
and
\begin{eqnarray}
&&q^B_+=\frac{1}{2}\frac{(\alpha\beta)^2g_A(\omega_+)g_B(\omega_+)\omega_+}
{\alpha^2g_A(\omega_+)T_A+
\beta^2g_B(\omega_+)T_B}
[T_B-T_A],
\label{curBBp}
\\
&&q^B_-=\frac{1}{2}\frac{(\alpha\beta)^2g_A(\omega_-)g_B(\omega_-)\omega_-}
{\beta^2g_A(\omega_-)T_A+\alpha^2g_B(\omega_-)T_B}
[T_B-T_A].
\label{curBBm}
\end{eqnarray}
At low temperatures $T_A, T_B\ll\omega_+,\omega_-$,
\begin{eqnarray}
&&q^A_+=\frac{(\alpha\beta)^2g_A(\omega_+)g_B(\omega_+)\omega_+}
{\alpha^2g_A(\omega_+)+\beta^2g_B(\omega_+)}[e^{-\omega_+/T_A}-e^{-\omega_+/T_B}],
\label{curAAAp}
\\
&&q^A_-=\frac{(\alpha\beta)^2g_A(\omega_-)g_B(\omega_-)\omega_-}
{\beta^2g_A(\omega_-)+\alpha^2g_B(\omega_-)}
[e^{-\omega_-/T_A}-e^{-\omega_-/T_B}].
\label{curAAAm}
\end{eqnarray}
and
\begin{eqnarray}
&&q^B_+=\frac{(\alpha\beta)^2g_A(\omega_+)g_B(\omega_+)\omega_+}
{\alpha^2g_A(\omega_+)+\beta^2g_B(\omega_+)}
\Big[e^{-\omega_+/T_B}-e^{-\omega_+/T_A}\Big],
\label{curBBBp}
\\
&&q^B_-=\frac{(\alpha\beta)^2g_A(\omega_-)g_B(\omega_-)\omega_-}
{\beta^2g_A(\omega_-)+\alpha^2g_B(\omega_-)}
\Big[e^{-\omega_-/T_B}-e^{-\omega_-/T_A}\Big].
\label{curBBBm}
\end{eqnarray}
In Fig.~\ref{fig4} we have depicted the currents $q_A$ and $q_B$ as a function of 
$T_A$ in the range $0.5$ to $1.5$ with $T_B=1$, with parameter choice 
$\omega_1=1.0$, $\omega_2=2.0$, $g_A(\omega_\pm)=g_b(\omega_\pm)=1$, 
and $\lambda=0.5$; note that the currents vanish in equilibrium for $T_A=T_B$.
\subsection{\label{redfield} Redfield approach}
In this section we discuss briefly the Redfield approach to the coupled qubit system.
Relaxing the delta function constraint or selection rule yielding the Lindblad equation
it follows from  Sec. \ref{general} that the kernel elements $K^{A,B}_{nn,23}, K^{A,B}_{nn,32},
K^{A,B}_{23,nn},K^{A,B}_{32,nn}$ for $n=1,4$ and $K^{A,B}_{23,23},K^{A,B}_{32,32}$ are
non vanishing indicating that the populations $\rho_{nn}$ couple to the coherences 
$\rho_{n\neq n'}$. The calculation is deferred to Appendix \ref{appendix}.

Considering the case $g_{A,B}(\omega_\pm)=g$ we find the populations and coherences
\begin{eqnarray}
\left(\begin{array}{c}
\rho_{11}\\\rho_{22}\\\rho_{33}\\\rho_{44}
\end{array}\right)
=
\frac{s^2+4e^2}{N}
\left(\begin{array}{c}
s _1s_2 \\s_1s_4 \\s_3s_2 \\s_3s_4
\end{array}\right)
-\frac{4k^2}{N}
\left(\begin{array}{c}
(s_1+s_4)(s_2+s_3)\\(s_1+s_4)(s_1+s_4)\\(s_2+s_3)(s_2+s_3) \\(s_1+s_4)(s_2+s_3)
\end{array}\right),
\label{rhod}
\end{eqnarray}
and
\begin{eqnarray}
\left(\begin{array}{c}
\rho_{23}\\\rho_{32}
\end{array}\right)
=
-\frac{2k}{N}
\left(\begin{array}{c}
(s_1s_2-s_3s_4)(s-2ie)\\(s_1s_2-s_3s_4)(s+2ie)
\end{array}\right).
\label{rhoc}
\end{eqnarray}
The trace condition $\sum_n\rho_{nn}=1$ yields the normalisation factor
\begin{eqnarray}
N=(s^2+4e^2)(s_1+s_3)(s_2+s_4)-4(ks)^2
\label{norm}
\end{eqnarray}
and we have, moreover, introduced the parameters
\begin{eqnarray}
&&s=s_1+s_2+s_3+s_4,
\label{s}
\\
&&e=E_{23}=-E_{32},
\label{E}
\\
&&k=
\frac{\alpha\beta g}{2}[n_A(\omega_+)-n_B(\omega_+)+n_A(\omega_-)-n_B(\omega_-)].
\label{k}
\end{eqnarray}
In equilibrium $T_A=T_B$  we have $k=0$. The populations are decoupled from
the coherences and the Redfield and Lindblad expression for the populations
are in agreement. In the non equilibrium case for $T_A\neq T_B$ we have $k\neq 0$
and the populations are reduced due to the coherences. The coherences $\rho_{23}$
and $\rho_{32}$ are complex conjugate (the density matrix is hermitian) and their
sign depends on $k\propto (n_A(\omega_\pm)-n_B(\omega_\pm))$, i.e., the sign
of $T_A-T_B$.

\begin{figure}.
\centering
\includegraphics[width=0.9\hsize]{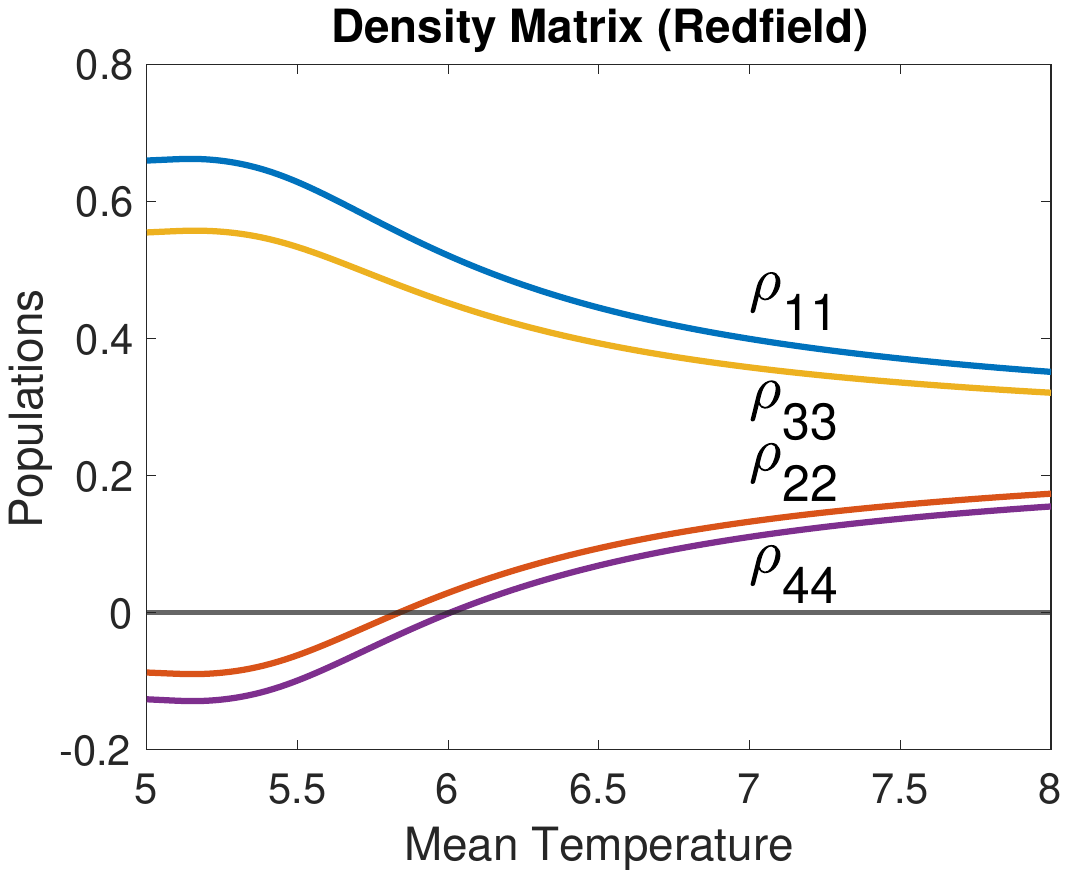}
\caption{
We depict the populations $\rho_{11}$, $\rho_{22}$, $\rho_{33}$, $\rho_{44}$ 
in the Redfield case as a function of the mean temperature $T$ for
$\Delta T=5$ The populations $\rho_{22}$ and $\rho_{44}$ are negative in 
the range $\Delta T<T\lesssim 6$.
}
\label{fig5}
\end{figure}

We have in Fig.~\ref{fig5} depicted 
the populations
$\rho_{nn}$ as a function of the mean temperature $T =(T_A+T_B)/2$
in the range $\Delta T<T<8$ with non equilibrium bias 
$\Delta T=(T_A-T_B)/2=5.0$ with parameter choice $\omega_1=1.0$, $\omega_2=2.0$, 
$g_A(\omega_\pm)=g_b(\omega_\pm)=1.0$, and $\lambda=0.5$. The plot shows that
$\rho_{44}$ and $\rho_{44}$ assumes negative value in the range $0<T\lesssim6.0$, 
indicating that the Redfield approach is inadequate in analysing the coupled qubit case.
\section{\label{discussion}Discussion and summary}
In this paper we have discussed heat currents in two qubit systems driven by two heat 
reservoirs maintained at different temperatures in order to test the validity of the second law
of thermodynamics in the microscopic domain, an important issue in quantum 
thermodynamics. The results for a single qubit system driven by two reservoirs are given in 
Sec. \ref{qubit} and are in accordance with the second law. The case of coupled qubits
is more involved, the results are presented in Sec. \ref{qubits} and are likewise in agreement
with the second law. We note that the current expressions in the single qubit and coupled
qubit cases have a similar structure. In deriving these results it is essential to apply a
global approach and operate in the basis of the full system Hamiltonian. Moreover, in order
to avoid an erroneous result it is also essential to use the Lindblad master equation
approach and in this way ensure a proper quantum channel with positive populations.

In order to discuss the possible role of coherences, we made use of a recent field
theoretical approach to open quantum system. This method yields at a first stage 
the Redfield master equation and has to be supplemented with a selection rule
in order to produce the Lindblad master equation. The Redfield equation applied
directly to the coupled qubit case gives rise to a coupling between coherences
and populations in the density matrix but yields negative populations and is
thus inadequate is describing the proper physics. 
\appendix*
\section{\label{appendix}The Redfield case}
Here we give details of the Redfield approach to the coupled qubit case. In the following
we assume identical reservoirs regarding their structure and define 
$g=g_{A,B}(\omega_\pm)$. We introduce the parameters $c_n, d_n$ according to the
scheme
\begin{eqnarray}
\left(\begin{array}{cc}
c_1 & d_1
\\
c_2 & d_2
\\
c_3 &d_3
\\
c_4 & d_4
\end{array}\right)
=\frac{\alpha\beta g}{2}
\left(\begin{array}{cc}
1+n_A(\omega_+) &1+n_B(\omega_+)
\\
1+n_A(\omega_-) & 1+n_B(\omega_-)
\\
n_A(\omega_+) &n_B(\omega_+)
\\
n_A(\omega_-) & n_B(\omega_-)
\end{array}\right).
\label{a1}
\end{eqnarray}
With the  kernel assigment
\begin{eqnarray}
K=
\left(
\begin{array}{cccccc}
K_{11,11}&K_{11,22}&K_{11,33}&0&K_{11,23}&K_{11,32}
\\
K_{22,11}&K_{22,22}&0&K_{22,44}&0&0
\\
K_{33,11}&0&K_{33,33}&K_{33,44}&0&0
\\
0&K_{44,22}&K_{44,33}&K_{44,44}&K_{44,23}&K_{44,32}
\\
K_{23,11}&0&0&K_{23,44}&K_{23,23}&0
\\
K_{32,11}&0&0&K_{32,44}&0&K_{32,32}
\end{array}
\right),
\label{a2}
\end{eqnarray}
and the abbreviations
\begin{eqnarray}
\left(\begin{array}{c}
 c_{12}
\\
 c_{34}
\\
d_{12}
\\
 d_{34}
\end{array}\right)
=
\left(\begin{array}{c}
 c_1+c_2
\\
 c_3+c_4
\\
 d_1+d_2
\\
  d_3+d_4
\end{array}\right),
\label{a3}
\end{eqnarray}
we find from the Redfield form in (\ref{Field-Kernel-Redfield}) the kernel 
$M_{pp',qq'}=-iE_{pp'}\delta_{pq}\delta_{p'q'}+K_{pp',qq'}$ for the determination
of the density matrix,
\begin{eqnarray}
M=
\left(
\begin{array}{cccccc}
-(s_3+s_4) &s_2&s_1 & 0&-k&-k
\\
s_4 &-(s_2+s_3) & 0 &s_1&0&0
\\
s_3 & 0 &-(s_1+s_4) & s_2&0&0
\\
0 &s_3&s_4&-(s_1+s_2)&k&k
\\
-k&0&0&k&-s/2-ie&0
\\
-k&0&0&k&0&-s/2+ie
\end{array}
\right).
\label{a4}
\end{eqnarray}
Here
\begin{eqnarray}
&&k=c_{12}-d_{12}=c_{34}-d_{34},
\label{a5}
\\
&&e=E_{23}=-E_{32},
\label{a6}
\end{eqnarray}
and by insertion
\begin{eqnarray}
k=\frac{\alpha\beta g}{2}[n_A(\omega_+)-n_B(\omega_+)+n_A(\omega_-)-n_B(\omega_-)].
\label{a7}
\end{eqnarray}
The parameter $k$ plays an important role. In equilibrium for $T_A=T_B$ we have 
$n_A(\omega_\pm)=n_B(\omega_\pm)$ yielding $k=0$.

The density matrix $\rho_{nn'}$ is determined by
\begin{eqnarray}
\sum_{qq'}M_{pp',qq'}\rho_{qq'}=0.
\label{a8}
\end{eqnarray}
By inspection of the kernel $M$ we have $\det(M)=0$ implying a solution.
We obtain the  diagonal density matrix elements (populations)
\begin{eqnarray}
\left(\begin{array}{c}
\rho_{11}\\\rho_{22}\\\rho_{33}\\\rho_{44}
\end{array}\right)
=
\frac{s^2+4e^2}{N}
\left(\begin{array}{c}
s _1s_2 \\s_1s_4 \\s_2s_3 \\s_3s_4
\end{array}\right)
-\frac{4k^2}{N}
\left(\begin{array}{c}
(s_1+s_4)(s_2+s_3)\\(s_1+s_4)(s_1+s_4)\\(s_2+s_3)(s_2+s_3) \\(s_1+s_4)(s_2+s_3)
\end{array}\right),
\label{a9}
\end{eqnarray}
whereas the off diagonal density matrix elements (coherences) are given by
\begin{eqnarray}
\left(\begin{array}{c}
\rho_{23}\\\rho_{32}
\end{array}\right)
=
-\frac{2k}{N}
\left(\begin{array}{c}
(s_1s_2-s_3s_4)(s-2ie)\\(s_1s_2-s_3s_4)(s+2ie)
\end{array}\right),
\label{a10}
\end{eqnarray}
where the trace condition $\sum_n\rho_{nn}=1$ yields the normalization factor
\begin{eqnarray}
N=(s^2+4e^2)(s_1+s_3)(s_2+s_4)-4(ks)^2.
\label{a11}
\end{eqnarray}
\newpage
\newpage
\clearpage

\end{document}